\documentstyle[12pt,aps,epsfig]{revtex}

\begin{document}

\title{On the polarization effects in $(p,n)$ reactions \\ between
the $A=48$ isobarical states}
\author{V. I. Isakov}
\address{Petersburg Nuclear Physics Institute\\
Gatchina, 188300 St. Petersburg, Russia}
\maketitle

\vspace{1cm}
\begin{center} \bf A b s t r a c t \end{center}

{\small
Isotopical dependence of spin-orbit splitting discovered by us in
spectra of heavy nuclei close to doubly magic ones is checked in
polarization effects arising in charge exchange $(p,n)$ reaction
between the $A=48$ isobarical states.}

\vspace{1cm}

Basing on the analysis of existing experimental data in nuclei close
to doubly magic nuclides $^{208}$Pb and $^{132}$Sn and on different
theoretical approaches it was shown in \cite{isa20} that
for similar orbitals the neutron
spin-orbit splitting in $N>Z$ nuclei is larger than the corresponding
proton splitting. It  was also demonstrated in \cite{isa20} that
different theoretical approaches lead to a larger neutron as compared
to proton splittings of $1d$ and $1p$ orbits in $^{48}$Ca, where the
corresponding experimental data on energies of single particle levels
are incomplete due to strong fragmentation effects. It was shown
in particular in \cite{isa20}, that in terms of
phenomenological potential the spectra of single
particle states, including the spin-orbit splittings, may be reproduced
by the average potential of the form
\begin{eqnarray}
&& \hat U\left(r,\hat\sigma,\tau_3\right)=V_0\left(1+\frac12\beta
\frac{N-Z}A\tau_3\right) f(r) +V_{ls}\left(1+\frac12\beta_{ls}
\frac{N-Z}A\tau_3\right)\frac1{r}\frac{df}{dr}\hat{\bf l}\hat{\bf s}
+\frac{(1+\tau_3)}2 U_{Coul}\ , \nonumber\\
&& f(r)=\left[1+\exp\left(\frac{r-R}a\right)\right]^{-1},
\quad R=r_0A^{1/3}
\end{eqnarray}
with $V_0=-51.5$ MeV, $r_0=1.27$ fm, $V_{ls}=33.2\,\rm MeV\cdot fm^2$,
$a\approx0.6\,$fm, $\beta=1.39$ and $\beta_{ls}\sim-0.6$; $\tau_3=-1$ for
neutrons and $\tau_3=+1$ for protons. Introducing the quantities
$t_3=-\tau_3/2$, $T_3=(N-Z)/2$ and making in the spirit of \cite{lane62}
the substitution $T_3\cdot t_3\to \hat{\bf T}\cdot\hat{\bf t}$, where
$\hat{\bf T}$ and $\hat{\bf t}$ are isospin vector operators for the core and
nucleon, we obtain the nuclear part of potential (1) in the isotopic-invariant
form (Lane potential), suitable for description of both the diagonal in
$t_3$ (single particle spectra and elastic scattering) and non-diagonal
(($pn)$-reactions leading to isobaric analogue states) processes:
\begin{eqnarray}
\hat U = V_0\left(1-2\beta\frac{\hat{\bf T}\cdot\hat{\bf t}}A\right)
f(r)+V_{ls}\left(1-2\beta_{ls}\frac{\hat{\bf T}\cdot\hat{\bf t}}A\right)
\frac1{r} \frac{df}{dr} \hat{\bf l}\hat{\bf s} .
\end{eqnarray}
A spin-orbit term in a potential
leads to polarization effects in scattering.
We see from (2) that while the polarization in elastic scattering depends
on the parameter combination of the form $V_{ls}\left(1-\beta_{ls}
\frac{(N-Z)}A t_3\right)\approx V_{ls}$, similar effects
in charge-exchange
reactions with excitation of isoanalogue states are proportional
to $\beta_{ls}\cdot V_{ls}$, and are  thus  defined by the
isovector mean spin-orbit field parameter $\beta_{ls}$,
as the $V_{ls}$ parameter is well known. Thus we can check the
conclusions of Ref. \cite{isa20} concerning the $\beta_{ls}$
value and based on nuclear spectra using the
data from $(p,n)$ quasielastic scattering. One can find corresponding
information about the polarization effects in the $^{48}$Ca region 
in Ref. \cite{and86}, where the $^{48}$Ca$(p,n)$ $^{48}$Sc
reaction with polarized protons leading to the 
 0$^+$ (6.67 MeV) isoanalogue state was studied, but with
theoretical analysis based on microscopical approach for description of
nuclear structure and in terms of nucleon-nucleon amplitudes (DWIA). Here
we proceed in terms of the Lane model basing on spin-orbit parameters
defined in \cite{isa20} and using the Born approximation
for the description of scattering. Similar problems for other target
nuclei were also studied in this approach in \cite{gos76}.

It is well known that in the Born approximation  polarization  effects
arising from the spin-orbit potential
disappear \cite{land}. Thus to describe these effects one needs to
introduce an imaginary part (absorption) into the optical potential, that
really means the account of effects beyond the Born approach. We
must also include in the real and imaginary parts of the potential the
dependence on the incident  energy, which was rather high ($E=134\,$MeV)
in \cite{and86}. In \cite{site}, \cite{bohr}  the following
proposition in the case of volume absorption is presented for the
$V_0$ parameter:  $V_0=V_0'(1-0.0058\cdot E)$ with
$V'_0=-52\,$MeV, that is rather close to the value of $-51.5\,$MeV
obtained  by us in \cite{isa20}. In this case the
corresponding absorption term in the optical potential was proposed in
\cite{site} in the form of \, $i\cdot W_Vf(r)$ with
$W_V$(MeV)$=-3.3 \cdot (1+0.03\cdot E)$. Surface absorption is usually
given as \, $i \cdot W_S(df/dr)$. For small values of transferred
momentum (small angles) both variants of absorption must result in
similar descriptions of the scattering process. In the case of \,
$a \ll R$ this leads to $W_S\approx-(R/3)W_V$. So, as an absorption term we use
the combination of the form
\begin{eqnarray}
i\cdot W_V[\alpha - (1-\alpha) \frac{R}{3} \frac{d}{dr}]f(r)
\end{eqnarray}
with $0 \leq \alpha \leq 1$, that leads to polarization effects,
independent on $\alpha$ at small scattering angles, but strongly
dependent on $\alpha$ at large values of transferred momentum. Thus,
for the description of polarization effects we use the optical potential
of the form (2), but with
\begin{eqnarray}
V_o \rightarrow -51.5 \cdot (1-0.0058 \cdot E) - i \cdot 3.3 \cdot
(1 + 0.03 \cdot E)[\alpha -
(1-\alpha) \frac{R}{3} \frac{d}{dr}],
\end{eqnarray}
adopting similar energy dependences for isoscalar and isovector terms
of the central nuclear potential.

In Fig.1 one can see the results of our calculations for the analyzing
power $A$ in the case of $(p,n)$ reaction on $^{48}$Ca leading to the
isoanalogue state
\begin{eqnarray}
A_{th} = \frac{{d\sigma_{\uparrow \uparrow}}/{d\omega}-
{d\sigma_{\uparrow \downarrow}}/{d\omega}}
{{d\sigma_{\uparrow \uparrow}}/{d\omega}+
{d\sigma_{\uparrow \downarrow}}/{d\omega}}\,\,; |A| \leq 1
\end{eqnarray}
together with experimental data and results of microscopical
calculations from \cite{and86}. Here $\sigma_{\uparrow \uparrow}$
and $\sigma_{\uparrow \downarrow}$ are cross sections with the polarization
vector $\vec {\varepsilon}$ of protons parallel or antiparallel to
$[\vec {k}_i \times \vec {k}_f]$.
We see that in the case of surface
absorption $(\alpha=0)$
our calculations that use the corresponding spin-orbit parameters from
\cite{isa20} demonstrate  good agreement with the experiment up
to high values of the scattering angle. At the same time, introduction
of the energy dependence into the spin-orbit parameter $V_{ls}$,
analogous to that for the central nuclear field, leads to poor
agreement with the experiment on the analyzing power.

Our calculations with $\alpha=0$ give the magnitude of differential cross
section for the $^{48}$Ca$(p,n)$ $^{48}$Sc$^*$ (I.A.S.) reaction on
unpolarized protons at zero angle equal to $\approx 7.7\,$mb/sr,
very weakly increasing with increase of
the parameter "$\alpha$", this cross section
sharply diminishes with the increase of the scattering angle and has some
structure at  $\Theta_{c.m.} \sim 20^0$. The value presented above
may be compared with the magnitude of  cross section at zero angle
measured in \cite{and85} ($\sim 7$ mb/sr), as
well as with the theoretical prediction \cite{and85} based on microscopical
theory ($\sim$ 7.5 mb/sr).

The following conclusions should be made:
\begin{itemize}
\item Experimental data on the isotopical dependence of spin-orbit
splitting in nuclei are consistent with the data on polarization effects
in $(p,n)$ quasielastic scattering. The mean field parameters defined
in \cite{isa20} that describe the proton and neutron spin-orbit splittings in
nuclei close to $^{132}$Sn and $^{208}$Pb, in particular the $\beta_{ls}$ one, 
well reproduce experimental data for $(p,n)$
quasielastic scattering on $^{48}$Ca.
\item Good description of analyzing power at high energy of incident
protons with the spin-orbit parameters borrowed from low energy spectroscopy
is consistent with supposition about the weak energy dependence of the
mentioned optical model parameters.
\item Satisfactory description of cross section for the $(p,n)$
reaction leading to isoanalogue state points to the correct
parameterization of the energy dependence of isovector terms in the
central nuclear potential used by us.
\item The obtained results unambiguously demonstrate a considerable
contribution of the surface absorption $((1-\alpha) \geq \,\sim 0.5)$ in
nuclei.
\end{itemize}

The author is grateful to B. Fogelberg, H. Mach, V.E. Bunakov and 
K.A. Mezilev for numerous and
useful discussions on the problems of spin-orbit splitting in nuclei.

This work was supported by the Russian Foundation of Fundamental Research
(grant No.00-15-96610).

\begin{figure}[h]
\centerline{\epsfig{file=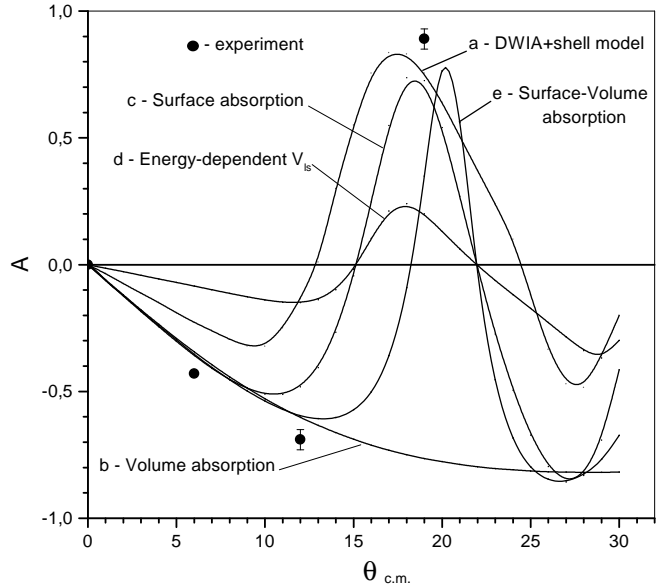,width=0.7\textwidth}}
\caption{\footnotesize
Experimental data on analyzing power [3] together with
results of different calculations:
a)~DWIA microscopical calculation [3].
b)~Our calculation with $\alpha=1$ (volume absorption),
$V_{ls}=33.2 \,\,MeV \cdot fm^2$, $\beta_{ls}=-0.6$.
c)~Our calculation with $\alpha = 0$ (surface absorption),
$V_{ls}=33.2 \,\,MeV \cdot fm^2$, $\beta_{ls}=-0.6$.
d)~Our calculation with $\alpha=0$, $\beta_{ls}=-0.6$,
and energy-dependent parameter $V_{ls}$.
e)~The same as "b", "c", but with $\alpha = 0.5$.
}
\end{figure}


\end{document}